\documentclass[aip,jcp,reprint]{revtex4-1}

\usepackage{graphicx} 
\usepackage{bm}
\usepackage{amsmath}
\usepackage{amssymb}

\providecommand{\eqnref}[1]{Eq.\ \eqref{#1}}

\providecommand{\mat}[1]{\ensuremath{\bm{\mathsf{#1}}}}
\DeclareMathOperator{\tr}{Tr}
\providecommand{\trans}[1]{\ensuremath{#1^{\intercal}}}
\renewcommand{\vec}[1]{\ensuremath{\mathbf{#1}}}
\def\<{\langle}
\def\>{\rangle}

\begin{document}

\title{Configuration-averaged 4f orbitals in ab initio calculations of
low-lying crystal field levels in lanthanide(III) complexes}

\author{Willem Van den Heuvel}
\author{Simone Calvello}
\author{Alessandro Soncini}
\email{asoncini@unimelb.edu.au}
\affiliation{School of Chemistry, The University of Melbourne, VIC 3010,
Australia}
\date{\today}

\begin{abstract}
A successful and commonly used {\it ab initio} method for the calculation of
crystal field levels and magnetic anisotropy of lanthanide complexes consists
of spin-adapted state-averaged CASSCF calculations followed by state
interaction with spin--orbit coupling (SI--SO). Based on two observations valid
for Ln(III) complexes, namely: (i) CASSCF 4f orbitals are expected to change
very little when optimized for different states belonging to the 4f electronic
configuration, (ii) due to strong spin--orbit coupling the total spin is not a
good quantum number, we show here via a straightforward analysis and direct
calculation that the CASSCF/SI--SO method can be simplified to a single
configuration-averaged HF calculation and one complete active space CI
diagonalization, including spin--orbit coupling, on determinant basis.  Besides
its conceptual simplicity, this approach has the advantage that all spin states
of the 4f$^n$ configuration are automatically included in the SO coupling,
thereby overcoming one of the computational limitations of the existing
CASSCF/SI--SO approach. As an example, we consider three isostructural complexes
[Ln(acac)$_3$(H$_2$O)$_2$], Ln = Dy$^{3+}$, Ho$^{3+}$, Er$^{3+}$, and find that
the proposed simplified method yields crystal field levels and magnetic
g-tensors that are in very good agreement with those obtained with
CASSCF/SI--SO.
\end{abstract}

\maketitle

\section{Introduction}

The ability of single-molecule magnets (SMMs) to display long-lived
spin-polarized states, which are of interest for the development of molecular
magnetic memories, is fundamentally rooted in the details of the electronic
structure of the ground state and the first few excited states of open-shell
metal complexes.\cite{librogiallo}  In particular, complexes of trivalent
lanthanide ions have recently proven to be promising to achieve SMM behavior at
higher temperatures than their transition metal
analogues.\cite{Ishikawa2003,SessoliPowell2009,Winpenny2013}  One
characteristic feature of the electronic structure of lanthanide complexes is
that the lowest energy electronic states do not differ significantly from
purely ionic states, leading to wave functions that are dominated by
spin--orbit atomic $J$-multiplets weakly split by the crystal field potential
of the surrounding ligands. This leads to unquenched 4f orbital angular
momentum in the ground state, making the magnetism of the open shell 4f
electrons particularly sensitive to the surrounding low-symmetry electrostatic
environment via strong spin--orbit coupling, therefore often strongly
anisotropic.  The resulting magnetic anisotropy in turn can lead to a high
spin-reorientation energy barrier in these complexes, which is at the origin of
slow magnetic relaxation dynamics and SMM behavior.\cite{librogiallo}

{\it Ab initio} calculations have proven very useful to help unravel the
crucial magneto-structural correlations that characterize new or potential
lanthanide SMMs.\cite{Luzon2012,Ungur2015,Liddle2015} The only {\it ab initio} method
currently used for this purpose is a combination of the complete active space
self consistent field method and the state interaction with spin--orbit
coupling method,\cite{Ungur2015} which usually goes by the acronym
CASSCF/RASSI--SO in the literature, after its implementation in the {\sc
Molcas} software package.\cite{Malmqvist2002,molcas8} We shall use the shorter
name CASSCF/SI--SO in this paper.

One of the first successful applications of the CASSCF\slash SI--SO method to
the magnetism of Ln(III) complexes was the explanation of a non-magnetic ground
state Kramers doublet in a triangular Dy(III) cluster.\cite{Toroidal2008} The
calculations predicted that the local magnetic anisotropy of the Dy centers
could be described by Ising-type spins whose local axes are tangential to the
triangle, resulting in a net cancellation of the total magnetic moment.
Subsequent studies used angle-dependent magnetic susceptibility measurements
on single crystals to provide direct evidence that such calculations are indeed
able to predict the direction of the magnetic easy axes and associated
g-factors in a number of low-symmetry Ln(III)
complexes.\cite{Bernot2009,Bernot2009a,Dota2012,Boulon2013,Jung2014,Boulon2013a}
Besides information on magnetic anisotropy, these {\it ab initio} calculations
also provide crystal field energy levels which can be compared with
experimental values, accessible through spectroscopic
techniques.\cite{Marx2014,Vonci2016}  CASSCF/SI--SO calculations have also
been used to rationalize the direction and extent of the magnetic
anisotropy in terms of ligand geometry and crystal field models based on
electrostatic charge
distributions,\cite{Chibotaru2012,Aravena2013,Chilton2013,Jung2015,Zhang2016}
to guide the design of new lanthanide-SMM candidates,\cite{Chilton2015}
and to investigate spin and orbital magnetization densities in a
series of lanthanide sandwich complexes.\cite{Gendron2015}

CASSCF/SI--SO is now widely used to compute spectroscopic and magnetic
properties of Ln(III) complexes. Here we present a critical assessment of this
approach and propose an alternative which is both a simplification and
extension. Section~\ref{casscf} reviews how CASSCF/SI--SO is applied to Ln(III)
complexes. A discussion of the characteristic electronic structure of 4f
elements suggests that state-dependent orbital flexibility, provided by CASSCF,
is of minor importance. This is corroborated by an analysis of the common
practice of applying state-averaged CASSCF to these systems. Based on these
findings we propose a simplified method based on one set of molecular orbitals,
obtained from a configuration-averaged Hartree-Fock (CAHF) calculation. This
approach allows a great simplification of the subsequent state-interaction
problem, which can now be formulated as a simultaneous diagonalization of
Coulomb repulsion and spin--orbit coupling in the basis of Slater determinants
of the 4f$^n$ configuration (CASCI--SO). 

In order to test the CAHF/CASCI--SO approach we apply it to three example
complexes. Some technical details of the method are described in Section
\ref{method}. Section \ref{results} compares the results with those obtained
with CASSCF/SI--SO and it is shown that a very good agreement is found.

\section{CASSCF/SI--SO treatment of the electronic structure
of Ln(III) complexes}\label{casscf}

We begin by reviewing the CASSCF/SI--SO method as it is applied to mononuclear
Ln(III) complexes in the molecular magnetism literature. (See, for example,
Ref.~\cite{Ungur2015} and references therein.) 

The purpose of the CASSCF step is to obtain wave functions that can be thought
of as corresponding to the atomic Russell--Saunders terms, whose degeneracies
are weakly split by the presence of the ligand environment. For a Ln(III)
complex whose formal configuration is 4f$^n$, this is achieved by choosing the
active space to consist of $n$ electrons in the seven 4f-like orbitals, giving
rise to $\binom{14}{n}$ Slater determinants. These are spin-adapted into
configurational state functions (CSF) of definite spin quantum numbers $S$.
For each spin manifold a number of CASSCF wave functions is then optimized.
Spin--orbit coupling (SOC) is introduced in the second step (SI--SO) by
diagonalizing the SOC operator in the basis of the optimized CASSCF wave
functions. The resulting eigenvectors are then used to calculate expectation
values of relevant operators, in particular the magnetic moment.

Due to the importance of SOC in the rare-earth coupling scheme known from
atomic theory, according to which the strengths of terms in the Hamiltonian are
ordered as follows: interelectronic repulsion  $>$ spin--orbit coupling $>$
crystal field potential, one would preferably include as many optimized CASSCF
spin states as possible in the SOC diagonalization.  The best possible
calculation in this setting would indeed include {\it all} spin states of the
4f$^n$ manifold in the spin--orbit mixing, corresponding to what in atomic
theory is known as complete intermediate coupling.\cite{CondonShortley} 

Note that CASSCF in general employs different molecular orbitals for different
states. Calculating a matrix element between any two such states can be
computationally expensive in large basis sets, due to the mutual
non-orthogonality of the molecular orbitals.\cite{Malmqvist1989} The RASSI
routine\cite{Malmqvist2002,molcas8} of {\sc Molcas} deals efficiently with this
problem, but it is still a computationally demanding task when interaction
between a large number of CASSCF states is required. For example, Dy(III), the
most studied lanthanide in single-molecule magnetism, has a 4f$^9$
configuration which corresponds to a total of 735 spin states. A complete
intermediate coupling calculation would require interaction between 735 CASSCF
states. This has so far not been feasible. In fact, in a recent review
concerning the application of CASSCF/SI--SO to Ln(III) complexes the authors
state that, based on experience, current computer capacities limit the number
of states to about 300.\cite{Ungur2015}

While CASSCF in principle allows individual optimization of CI roots with
respect to molecular orbital rotations, this is in practice only feasible for a
few of the lowest-energy roots of the CI matrix. When a large number of roots
is required, the only viable strategy is to resort to state-averaged CASSCF,
whereby the molecular orbitals are optimized to minimize the average energy of
the required roots.  Thus, within the current application of the CASSCF/SI--SO
strategy to Ln(III) complexes, for each $S$, a state-averaged CASSCF
calculation is performed often including \emph{all} states with spin $S$ that
are possible within the 4f$^n$ manifold, giving equal weight to all states in
the average.  It should be noted that the molecular orbitals so obtained are
completely independent from the CI problem. This can be seen as follows.  The
CASSCF iterative process involves alternating orbital rotation and CI
diagonalization steps. The latter determines the roots and their energy. But if
only the average energy of \emph{all} roots is required, then diagonalization
is not necessary, because the sum of all eigenvalues is always equal to the sum
of all diagonal matrix elements (i.e.\ the trace of the CI matrix):
\begin{equation}\label{eq:trace} 
\sum_{i=1}^{\dim\mat{H}_S} E_{S,i} = \tr \mat{H}_S.  
\end{equation}
A CASSCF optimization of this type is thus mathematically equivalent to a
minimization of the trace of the CI Hamiltonian matrix with respect to
molecular orbital rotations. Since the trace is independent of basis choice,
there is no need to build and diagonalize the CI matrix at every iteration.
Indeed, one obtains exactly identical results by first performing an SCF
minimization of $\tr \mat{H}_S$ and then, using the orbitals so obtained, a
single CI diagonalization to obtain state energies and wave functions. This
shows that the orbital optimization is completely decoupled from the CI problem. 

The reason that this state-averaging procedure works well is probably that
molecular orbital relaxation between states of the 4f$^n$ space is relatively
small. In fact, the almost pure atomic nature of the 4f valence shell would
suggest that MCSCF calculations are not required.  Characteristic for most
Ln(III) complexes is indeed the almost complete absence of covalent mixing of
4f atomic orbitals with ligand orbitals. It can therefore be argued that the
strong electron correlation problem is essentially an atomic one. In atoms
Coulomb repulsion commutes with both $\bm{S}^2$ and $\bm{L}^2$ electronic
angular momenta, so that if one could start with CSF's with good quantum
numbers $S$ and $L$ arising from a given 4f$^n$ configuration, i.e.\ a
representation of the Russell--Saunders terms, one would expect such basis to
be optimal in order to capture the dominant features of strong electron
correlation (static correlation) at the atomic level, or in a few cases after
solving very small diagonalization problems for the determination of accurate
Russell--Saunders terms as linear combinations of few $LS$-symmetry adapted
CSF's. It is therefore arguable that pure atomic spin--orbit multiplets
characterized by a total angular momentum quantum number $J$, obtained by
diagonalization of the SOC Hamiltonian on the basis of atomic Russell--Saunders
CSF's, would represent the most appropriate guess-states for subsequent
molecular calculations, possibly for simple SCF calculations to determine the
optimal orbitals in the presence of the crystal field potential. However,
current molecular quantum chemistry codes do not work that way and one is
typically forced to start from a basis of CSF's that are only spin-symmetry
adapted, thus quite far from being atomic states already taking care of on-site
correlation. 

If this argument is valid, and we are only interested in the properties of the
lowest spin-orbit multiplet, we should expect that in the CASSCF/SI--SO
approach what captures the relevant electron correlation effects is the attempt
to reproduce $\bm{L}^2$ eigenfunctions as closely as the rotational
symmetry-breaking effect of the ligands allows, pursued via diagonalization of
Coulomb repulsion in the basis of the CAS same-spin CSF's. Such attempt would
arguably be quite independent of orbital optimization which is instead crucial
to represent the symmetry-breaking character of the crystal field electrostatic
potential, and consequent splitting of the ground atomic multiplet.

The previous discussion naturally suggests  to go one step further and assume
that the averaged 4f orbitals will not depend much on the total spin of the
wave function either, so that just one set of orbitals can be used to describe
the entire 4f$^n$ manifold of states. The result of this simplification is that
we can now formulate the CI on a determinantal basis and that the CI on the
electrostatic Hamiltonian and on the SOC, which are separated in CASSCF/SI--SO,
can now be combined in just one diagonalization step. This approach will be
described in the next section.

\section{Simplified approach: CAHF/CASCI--SO} \label{method}

The method we propose here consists of two steps. In the first step a set of
optimal molecular orbitals is obtained from a suitable configuration-averaged
restricted HF-type calculation as detailed in subsection~\ref{subCAHF}.  In the
second step the optimized orbitals are used to construct {\it all} Slater
determinants of the open 4f$^n$ shell, regardless of which subset was used in
the first step to build the average energy functional.  These determinants form
the basis for a generalized configuration-interaction type matrix
diagonalization which, besides the usual non-relativistic Coulomb repulsion
operator, also involves the spin--orbit coupling operator, as detailed in
subsection~\ref{subSocasci}.

Configurational-average methodologies have been extensively discussed and
applied for many years, since the work of McWeeny who proposed them to treat
excited states associated with any number of open
shells.~\cite{McWeeny1974,McWeeny1989} We also note in passing that, as a
cheaper alternative to CASSCF, complete active space configuration interaction
(CASCI) based on molecular orbitals determined in a previous step has also been
studied by several workers in a variety of
contexts.\cite{Bofill1989,Potts2001,Granucci2001,Abrams2004,Slavicek2010,Shu2013,Shu2015,Keller2015}
Visser et al.\cite{Visser1992} applied the idea of configuration-averaged
orbitals to a lanthanide crystal-impurity problem in the context of
relativistic four-components calculations, which therefore contain spin--orbit
coupling from the very start.  Their method has never developed, to our
knowledge, into a practical non-relativistic \textit{ab initio} approach
dedicated to the calculation of crystal field levels and magnetic properties of
lanthanide complexes.

\subsection{Hamiltonian} 
\label{subHam}

The Hamiltonian that we use in the present paper is identical to that used for
CASSCF/SI--SO calculations in \textsc{Molcas}.\cite{molcas8} It is given by the
second order Douglas--Kroll--Hess (DKH) scalar Hamiltonian, combined with the
usual non-relativistic Coulomb electron repulsion operator and an effective
one-electron atomic mean-field (AMFI) approximation of the DKH no-pair
spin--orbit operator.\cite{Hess1996, Malmqvist2002} A detailed description of
this Hamiltonian can be found in Ref.~\cite{Roos2004}.

We have implemented configuration-averaged Hartree--Fock (CAHF) and
spin--orbit-inclusive complete-active space configuration interaction
(CASCI--SO) modules in a local development code, \textsc{Ceres} (\textit{Computational
Emulator of Rare Earth Systems}), which is based
on the open-shell version~\cite{Soncini2007} of the \textsc{Sysmo}
software.\cite{SYSMO} Since in \textsc{Sysmo} the integrals of the DKH
operators in the atomic basis set are not available, all integrals are computed
using the SEWARD program of \textsc{Molcas 8.0},\cite{molcas8} and read into
\textsc{Ceres} to be used by our CAHF and  CASCI--SO modules, which are
described in the next sections.

\subsection{Configuration-averaged Hartree--Fock (CAHF) orbitals}
\label{subCAHF}

In order to obtain a set of molecular orbitals (MO) for the subsequent CI
calculation we minimize the average energy of the states of the 4f$^n$
configuration.\cite{McWeeny1974} There are several ways one might choose to do
this. For instance, we could build an average-energy functional over all Slater
determinants with a constant $M_S$ projection of the spin angular momentum.
Such approach turns out to be rather useful, e.g. for debugging purposes. In
fact, when $M_S = S_{\mathrm{max}}$, the energy functional is equivalent to
that minimized during a state-averaged CASSCF calculation performed with as
many roots as there are $S=S_{\mathrm{max}}$ states within the 4f$^n$
configuration (see \eqnref{eq:trace}), and thus the orbitals obtained via this
approach should be equivalent to those obtained via a state-averaged CASSCF
calculation with \textsc{Molcas}. Moreover, for any other $M_S$ value our
configuration-averaged approach optimizes orbitals by mixing two or more
spin-manifolds. Clearly, such orbitals are not equivalent to any of the
state-averaged CASSCF optimized orbitals, and it is thus an interesting
question whether this can lead to significant discrepancies between the two
approaches. Finally, we can average over all $M_S$ spin manifolds and simply
obtain a fully-averaged SCF problem within the 4f$^n$ configuration, in the
spirit of the old McWeeny proposal,\cite{McWeeny1974} which, if our reasoning
is correct, should yet again lead to nearly atomic 4f-like orbitals with no
appreciable discrepancies from the SA--CASSCF methodology.

McWeeny has treated the general case of configurational averaging (see, for
example, Ref.~\cite{McWeeny1989}, \S\S 6.5--6.6). We briefly present here
the special case of $M_S$-configuration-averaging, and then generalize the
results to those reported by McWeeny. We use McWeeny's density matrix notation.

Let us consider a system with one closed and one open shell, having $n_1$
spatial orbitals in the closed shell (or $n_1$ inactive orbitals), $n_2$
spatial orbitals in the open shell (or $n_2$ active orbitals). Let $n_A$ be the
number of active electrons, with $n_{\alpha}$ spin-up and $n_{\beta}$ spin-down
electrons ($n_A = n_{\alpha}+n_{\beta}$ ), thus with fixed
$M_S=1/2(n_{\alpha}-n_{\beta})$ value. Averaging the electrostatic Hamiltonian
over the $\binom{n_2}{n_{\alpha}}\binom{n_2}{n_{\beta}}$ Slater determinants
that can be formed by occupying the $n_2$ active orbitals with $n_{\alpha}$
spin-up electrons and $n_{\beta}$ spin-down electrons we obtain the energy
functional: 
\begin{subequations}\label{eq:EavMS} \begin{align}
        E_\text{av}^{M_S}&= \nu_1 \sum_{i}^{n_1} h_i + \frac{\nu_1^2}{2}
        \sum_{i,j}^{n_1}(J_{ij}-\frac{1}{2}K_{ij}) \label{eq:Eav_a}\\
        &+\nu_1\nu_2\sum_i^{n_1}\sum_u^{n_2}(J_{iu}-\frac{1}{2}K_{iu})\label{eq:Eav_b}\\
        &+\nu_2 \sum_{u}^{n_2}h_u + \frac{\nu_2^2}{2}  \sum_{u,v}^{n_2}
        (\lambda_J^{M_S} J_{uv}- \lambda_K^{M_S} K_{uv}). \label{eq:Eav_c}
\end{align} 
\end{subequations} 
where $\nu_1= 2$, $\nu_2= \frac{n_A}{n_2}$ are
the (average) occupations of the two sub-shells, indices $i,j$ run over the
inactive space, indices $u,v$ run over the active space, and the Coulomb
($J_{uv}$) and exchange ($K_{uv}$) integrals for the active space are weighted
by the $M_S$-dependent coefficients: 
\begin{equation} \label{msaverage}
        \begin{split} \lambda_J^{M_S}& = \frac{n_2 n_A \left(n_A -1\right)- 2
n_{\alpha}n_{\beta} }{n_A^2 \left( n_2-1\right)} \\ 
\lambda_K^{M_S} &=
\frac{n_2 n_A \left(n_A -1\right)- 2 n_2 n_{\alpha}n_{\beta} }{n_A^2 \left(
n_2-1\right)} \end{split} 
\end{equation} 

This expression can be rewritten in terms of density and integral matrices on
the atomic basis 
\begin{equation}\label{eq:EavMSdensity}
        E_\text{av} = \nu_1\tr[ \mat{R}_1(\mat{h}+\tfrac{1}{2}\mat{G}_1)] + 
        \nu_2\tr [ \mat{R}_2(\mat{h}+\tfrac{1}{2}\mat{G}_2)],
\end{equation}
where  $\mat{R}_i=\mat{T}_i\trans{\mat{T}_i}$ is the
density matrix of shell $i$, whose LCAO coefficients are contained in the columns
of $\mat{T}_i$, \mat{h} is the matrix of the one-electron Hamiltonian, and 
\begin{equation}
\label{g1g2}
        \begin{split}   
             \mat{G}_1 & = \nu_1\mat{G}(\mat{R}_1) +\nu_2\mat{G}(\mat{R}_2)\\
             \mat{G}_2 & = \nu_1\mat{G}(\mat{R}_1) +\nu_2\mat{G}^{M_S}(\mat{R}_2)
        \end{split}
\end{equation}
are Coulomb--exchange matrices, with
\begin{equation}
\begin{split}
        \mat{G}(\mat{R}) & =  \mat{J}(\mat{R})-\tfrac{1}{2}\mat{K}(\mat{R})\\
        \mat{G}^{M_S}(\mat{R}) & =  \lambda_J^{M_S} \mat{J}(\mat{R})- \lambda_K^{M_S} \mat{K}(\mat{R}),
\end{split}
\end{equation}
and
\begin{equation}
        \begin{split}
              \mat{J}(\mat{R})_{\alpha\beta}&=\sum_{\delta\gamma}
              \mat{R}_{\delta\gamma}\langle\gamma\alpha|\delta\beta \rangle \\
              \mat{K}(\mat{R})_{\alpha\beta}&=\sum_{\delta\gamma}
              \mat{R}_{\delta\gamma}\langle\gamma\alpha|\beta\delta\rangle
      \end{split}
\end{equation}
the usual Coulomb and exchange matrices, respectively. Greek letters denote
basis set functions. 

We are also interested in optimizing the energy functional originally proposed
by McWeeny, arising from averaging over the full set of Slater determinants
that can be obtained in the chosen active space, regardless of the
spin-projection quantum number $M_S$. This can be easily recovered by
multiplying $\lambda_J^{M_S}$ and $\lambda_K^{M_S}$ in Eqs.~(\ref{msaverage})
by the number of Slater determinants
$\binom{n_2}{n_A/2+M_S}\binom{n_2}{n_A/2-M_S}$
 with constant $M_S$, summing over all possible $M_S$, and dividing
by the total number of Slater determinants. This leads to:
\begin{align} 
\label{fullaverage} 
\begin{split} 
\overline{\lambda}_J &=    \frac{\sum\limits_{M_S}
\dbinom{n_2}{\frac{n_A}{2}+M_S}\dbinom{n_2}{\frac{n_A}{2}-M_S}\lambda_J^{M_S}}
{\sum\limits_{M_S}
\dbinom{n_2}{\frac{n_A}{2}+M_S}\dbinom{n_2}{\frac{n_A}{2}-M_S}} =
\frac{2}{\nu_2}\frac{\left(n_A-1\right)}{\left(2 n_2-1\right)} \\
\overline{\lambda}_K &= \frac{\sum\limits_{M_S}
\dbinom{n_2}{\frac{n_A}{2}+M_S}\dbinom{n_2}{\frac{n_A}{2}-M_S}\lambda_K^{M_S} }
{\sum\limits_{M_S}
\dbinom{n_2}{\frac{n_A}{2}+M_S}\dbinom{n_2}{\frac{n_A}{2}-M_S} } =
\frac{1}{\nu_2}\frac{\left(n_A-1\right)}{\left(2 n_2-1\right)} 
\end{split} 
\end{align}
which determines a modified energy functional, given by a modified
Eq.~(\ref{eq:EavMS}), where the average electron--electron repulsion term within
the active space (i.e., last term in the last line in Eq.~(\ref{eq:EavMS})) is
modified using  Eq.~(\ref{fullaverage}), thus becoming: 
\begin{equation}\label{mcweenyresult} 
  \nu_2\frac{\left(n_A-1\right)}{\left(2 n_2-1\right)}
\sum_{u,v}^{n_2} (J_{uv}- \frac{1}{2} K_{uv}).  
\end{equation}

The energy functional Eq.~(\ref{eq:EavMS}), or equivalently that originally
proposed by McWeeny that can be obtained from Eq.~(\ref{mcweenyresult}), is
akin to the energy functional arising in the restricted open-shell HF (ROHF)
theory.\cite{McWeeny1989}  Thus it can be easily shown that a sufficient
condition to minimize Eq.~(\ref{eq:EavMS}) or Eq.~(\ref{eq:EavMSdensity}) 
is to build the density matrices for inactive and active spaces from the
converged self-consistent eigenfunctions of an effective Fock-like Hamiltonian:
\begin{equation}
\label{fock2}
\mat{F}_{\mathrm{eff}} = a \tilde{\mat{R}}_2 \mat{F}_1  \tilde{\mat{R}}_2 
+ b \tilde{\mat{R}}_1  \mat{F}_2  \tilde{\mat{R}}_1 
+ c  \tilde{\mat{R}}_3 \left(\nu_1 \mat{F}_1 - \nu_2 \mat{F}_2\right)  \tilde{\mat{R}}_3
\end{equation}
where $\mat{F}_i = \mat{h}+\mat{G}_i$,  $\tilde{\mat{R}}_i = \mat{1} -
\mat{R}_i$, and $a$, $b$ and $c$ are arbitrary real non-zero parameters that
can be adjusted to improve convergence of the SCF process.  This procedure has
been implemented via a simple modification of the open-shell \textsc{Sysmo}
code~\cite{Soncini2007}, by modifying the mean-field repulsion potential within
the active space by Eqs.~(\ref{msaverage}) or Eqs.~(\ref{fullaverage}), which
enter Eq.~(\ref{fock2}) via the G-matrix $\mat{G}_2$ in Eq.~(\ref{g1g2}). For
the time being the only tools we have implemented to achieve convergence of the
configuration-averaged SCF process are a direct inversion of the iterative
subspace (DIIS) algorithm,\cite{Pulay1980} and level shifters for the open and
virtual shells.\cite{Guest1974}

\subsection{Spin--orbit-including complete active space configuration
interaction (CASCI--SO)}
\label{subSocasci}

The CAHF orbitals are now used to set up the full configuration interaction
calculation in the determinantal basis of the open shell. Since there are 7
spatial 4f orbitals and $n$ electrons distributed among them, the dimension of
the CI secular matrix is $\binom{14}{n}$. The largest dimension occurs for
$n=7$ and is 3432. As this is a relatively small number, diagonalization of the
CI matrix is fast.  

The CI routine that we have implemented is based on the $\sigma$-algorithm of
Olsen \textit{et al.}\cite{Olsen1988} for the scalar (spin-conserving) part of
the Hamiltonian (i.e., the scalar one-electron plus interelectronic Coulomb
repulsion terms). We also need to include the CI matrix elements of the SOC
operator, which is not spin-conserving.

The AMFI SOC operator can generally be written as 
\begin{equation} 
        H^\text{SO} = \sum_i \vec{t}(i)\cdot\vec{s}(i), 
\end{equation} 
where the summation is over all electrons, and $\vec{t}$ is a function of space
only. In second quantization this gives
\begin{multline}
        H^\text{SO} =\frac{1}{2}
        \sum_{u,v}[t^z_{uv}a^\dagger_{u\alpha}a_{v\alpha}-
                t^z_{uv}a^\dagger_{u\beta}a_{v\beta}
\\        +(t^x_{uv}-it^y_{uv})a^\dagger_{u\alpha}a_{v\beta}
+(t^x_{uv}+it^y_{uv})a^\dagger_{u\beta}a_{v\alpha}], 
\end{multline}
where $\vec{t}_{uv}$ are the AMFI integrals, transformed to the active
molecular orbital basis. The first two terms are spin-conserving, and can be
handled by the scalar CI algorithm.\cite{Olsen1988} The third and the fourth
term introduce spin flips and in order to include these we have supplemented the
original algorithm with routines that handle single excitations from $\alpha$
to $\beta$ spin orbitals and vice versa. 

\section{Application}\label{results}

This section presents results of calculations on three isostructural complexes:
[Ln(acac)$_3$(H$_2$O)$_2$], Ln = Dy, Ho, Er.\cite{Jiang2010} We compare the
CASSCF/SI--SO method with the CAHF/CASCI--SO method.

We performed single-point calculations on all three complexes, using the
crystallographic structures.\cite{Jiang2010} ANO-RCC basis sets\cite{Roos2008}
were used on all atoms, contracted to [9s8p6d4f3g2h] for Dy, Ho, Er, [3s2p1d]
for C and O, and [2s1p] for H. 

CASSCF/SI--SO calculations were done with \textsc{Molcas 8.0}.\cite{molcas8}
The active space consists of the seven Ln 4f orbitals, and is occupied by 9
electrons for Dy, 10 electrons for Ho, and 11 electrons for Er. These
occupations correspond to the trivalent oxidation state and give rise to a
ground spin--orbit multiplet with total angular momentum $J=15/2$ for Dy(III),
$J=8$ for Ho(III), and $J=15/2$ for Er(III). (Within the single
Russell--Saunders term approximation, these correspond to 
${^6}\mathrm{H}_{\frac{15}{2}}$ for Dy(III), ${^5}\mathrm{I}_{8}$ for Ho(III),
and ${^4}\mathrm{I}_{\frac{15}{2}}$ for Er(III)). Crystal field splitting of
the $J$ multiplets results in eight low-lying Kramers doublets (KD's) for the 
Dy and Er compounds. The ground $J$ multiplet of the Ho compound, being an
even-electron system, splits into seventeen non-degenerate states. State-averaged (SA)
CASSCF optimizations were done on the average energy of all states belonging to
the highest spin, \textit{viz}.\ 21 $S=5/2$ states for Dy, 35 $S=2$ states for
Ho, and 35 $S=3/2$ states for Er. The resulting wave functions were spin--orbit
coupled with the RASSI module of {\sc Molcas}. We chose not to include states
of lower spin in the spin--orbit coupling calculation for two reasons: First,
as mentioned in Section \ref{casscf}, it is computationally not feasible to
include all spins of the 4f$^9$ manifold of Dy(III) in a RASSI calculation
(this goes for the 4f$^{10}$ manifold of Ho(III) as well). In
practice, one resorts to an approximation, either by discarding states above a
certain cutoff energy, or by just including the highest-spin states
only.\cite{Ungur2015} Second, this allows to assess the influence of SOC mixing
with states of lower spin, by comparison with the CASCI--SO results.

CAHF/CASCI--SO calculations were done following the method described in Section
\ref{method}. Two different types of HF averaging were considered to obtain the
molecular orbitals: in the first, averaging was done over all determinants with
maximum spin projection $M_S$, using the $\lambda$ coefficients in
\eqnref{msaverage}. In the second type averaging was done over \emph{all}
determinants, using the $\overline{\lambda}$ coefficients in
\eqnref{fullaverage}. 

Magnetic g-factors were calculated for each Kramers doublet of the Dy and Er
complexes.\cite{Bolvin2006, Vancoillie2007} The Ho complex is an even-electron
system and as such has no Kramers doublets. Nevertheless, it is sometimes
possible to find two quasi-degenerate states and treat them as a
pseudo-doublet, for which g-factors can be calculated. This has been done for
the two lowest states of the Ho complex (Table \ref{tb:ho}). Note that such
pseudo-doublets have only one non-zero principal g-factor.\cite{Griffith1963}

Table \ref{tb:totenLn} presents calculated average and
ground state energies, the latter both with and without inclusion of SOC. Note
that the SOC-free energies in columns 1 and 2 are identical. This confirms the
equivalence of SA--CASSCF and CAHF orbitals for the high-spin subspace,
predicted by \eqnref{eq:trace}.

The calculated relative energies and magnetic g-factors are summarized in
Tables \ref{tb:dy}--\ref{tb:er}. It is clear that there is minimal difference
between the results generated by the three methods. The largest changes are
observed when including all spin states in the spin--orbit coupling, as opposed
to the high-spin states only. Smaller changes are observed when using orbitals
averaged over all states, as opposed to orbitals averaged over the high-spin
states only.

\begin{table}
\caption{[Ln(acac)$_3$(H$_2$O)$_2$]: Comparison of total energies, shifted by
$-13328$ Hartree for Dy, $-13787$ Hartree for Ho, and $-14256$ Hartree for Er. 
GS = ground state energy.\label{tb:totenLn}}        
\small
\begin{tabular*}{\columnwidth}{@{\extracolsep{\fill}}lllll}
\hline
&&SA--CASSCF & CAHF  &CAHF\\
Ion&&on $S_{\mathrm{max}}$ & on $M_S=S_{\mathrm{max}}$ & on all $M_S$\\
\hline
Dy&$E_\text{av}$ & $-1.021652$ & $-1.021652$ & $-0.777037$\\
&&       &       \multicolumn{2}{c}{CASCI}\\
\cline{4-5}
&GS  & $-1.057170$ & $-1.057170$ & $-1.055993$\\
&&  SI--SO     &   \multicolumn{2}{c}{CASCI--SO}\\
\cline{3-5}
&GS & $-1.079553$ & $-1.086042$ & $-1.084697$\\
\rule{0pt}{5ex}
Ho&$E_\text{av}$ & $-0.477551$ & $-0.477551$ & $-0.327231$\\
&&       &       \multicolumn{2}{c}{CASCI}\\
\cline{4-5}
&GS  & $-0.559519$ & $-0.559519$ & $-0.558906$\\
&&  SI--SO     &   \multicolumn{2}{c}{CASCI--SO}\\
\cline{3-5}
&GS & $-0.589286$ & $-0.595589$ & $-0.594871$\\
\rule{0pt}{5ex}
Er&$E_\text{av}$ & $-0.866271$ & $-0.866271$ & $-0.789330$\\
&&       &       \multicolumn{2}{c}{CASCI}\\
\cline{4-5}
&GS  & $-0.949870$ & $-0.949870$ & $-0.949656$\\
&&  SI--SO     &   \multicolumn{2}{c}{CASCI--SO}\\
\cline{3-5}
&GS & $-0.982882$ & $-0.986258$ & $-0.985998$\\
\end{tabular*}        
\end{table}

\begin{table}
\small
\caption{[Dy(acac)$_3$(H$_2$O)$_2$]: Calculated relative energies and
g-factors of the Kramers doublets corresponding to the crystal-field split $J=15/2$ 
ground multiplet. \label{tb:dy}}
\begin{tabular*}{\columnwidth}{@{\extracolsep{\fill}}lllll}
\hline
Doublet  & Energy/cm$^{-1}$ &$g_1$ &$g_2$&$g_3$\\
        \hline
      &\multicolumn{4}{c}{SA--CASSCF/SI--SO ($S=5/2$)}\\
        \hline
1&0.0   & 0.01 & 0.01 & 19.56 \\
2&156.4 & 0.26 & 0.45 & 15.70 \\
3&234.8 & 2.02 & 2.87 & 11.28 \\
4&289.5 & 2.27 & 5.87 & 7.01\\
5&323.3 & 2.12 & 4.25 & 13.84 \\
6&417.9 & 0.01 & 0.13 & 16.28 \\
7&477.7 & 0.04 & 0.08 & 18.84 \\
8&539.7 & 0.01 & 0.02 & 19.22 \\
\hline
&\multicolumn{4}{c}{CAHF ($M_S=5/2$)/CASCI--SO}\\
        \hline
1&0.0   & 0.01 & 0.01 & 19.44  \\
2&154.0 & 0.26 & 0.45 & 15.60 \\
3&231.9 & 1.92 & 2.75 & 11.22 \\
4&285.8 & 2.42 & 6.02 & 6.99 \\
5&319.1 & 2.05 & 4.13 & 13.66  \\
6&410.8 & 0.00 & 0.13 & 16.21 \\
7&468.9 & 0.04 & 0.07 & 18.78 \\
8&529.8 & 0.01 & 0.02 & 19.14 \\
\hline
&\multicolumn{4}{c}{CAHF (all $M_S$)/CASCI--SO}\\
        \hline
1&0.0   & 0.01 & 0.01 & 19.43 \\
2&155.0 & 0.25 & 0.43 & 15.59 \\
3&234.2 & 1.87 & 2.68 & 11.23 \\
4&288.5 & 2.40 & 6.05 & 6.97\\
5&321.2 & 2.05 & 4.22 & 13.60 \\
6&413.3 & 0.01 & 0.13 & 16.23 \\
7&471.8 & 0.04 & 0.07 & 18.79 \\
8&533.3 & 0.01 & 0.02 & 19.14 \\
\end{tabular*}
\end{table}

\begin{table}
\small
\caption{[Ho(acac)$_3$(H$_2$O)$_2$]: Calculated relative energies of the states
corresponding to the crystal-field split $J=8$ ground multiplet. The 
calculated g-factors are those of the pseudo-doublet consisting of states 1
and 2. \label{tb:ho}}
\begin{tabular*}{\columnwidth}{@{\extracolsep{\fill}}llll}
\hline
&SA--CASSCF/  & CAHF ($M_S=2$)/  &CAHF (all $M_S$)/\\
&SI--SO ($S=2$)&CASCI--SO        &CASCI--SO\\ 
\hline
$g_1$& 0.00 &0.00 &0.00\\
$g_2$& 0.00 &0.00 &0.00\\
$g_3$& 17.23&17.12&17.16\\
\\
State& Energy/cm$^{-1}$&&\\
%\rule{0pt}{3ex}\\
1&0.0&0.0&0.0 \\
2&4.2&4.0&3.9 \\
3&34.9&33.2&33.6\\
4&45.0&42.7&43.0\\
5&101.4&96.1&96.5\\
6&125.8&118.8&119.2\\
7&146.1&138.1&138.4\\
8&162.5&153.5&154.1\\
9&177.6&168.3&168.4\\
10&209.0&197.7&197.5\\
11&220.1&208.1&208.1\\
12&222.8&210.6&210.8\\
13&230.9&218.5&218.3\\
14&255.6&242.0&242.2\\
15&264.3&250.1&250.0\\
16&293.9&277.6&277.4\\
17&295.0&278.6&278.4\\
\end{tabular*}
\end{table}
\begin{table}
\small
\caption{[Er(acac)$_3$(H$_2$O)$_2$]: Calculated relative energies and
g-factors of the Kramers doublets corresponding to the crystal-field split
$J=15/2$ ground multiplet.\label{tb:er}}
\begin{tabular*}{\columnwidth}{@{\extracolsep{\fill}}lllll}
\hline
Doublet  & Energy/cm$^{-1}$ &$g_1$ &$g_2$&$g_3$\\
        \hline
      &\multicolumn{4}{c}{SA--CASSCF/SI--SO ($S=3/2$)}\\
        \hline
1&0.0 &  0.65 & 1.88 & 14.43 \\
2&29.6&  1.92 & 3.48 & 12.24 \\
3&70.4&  1.74 & 4.41 & 9.45 \\
4&87.5&  0.27 & 3.64 & 10.85 \\
5&135.3& 0.19 & 3.69 & 10.37 \\
6&179.1& 1.56 & 3.82 & 10.99 \\
7&252.2& 0.15 & 2.64 & 11.53 \\
8&296.6& 0.63 & 2.11 & 15.26 \\
\hline
&\multicolumn{4}{c}{CAHF ($M_S=3/2$)/CASCI--SO}\\
        \hline
1&0.0 & 0.57 & 1.68 & 14.62 \\
2&29.2& 1.81 & 3.21 & 12.59 \\
3&70.9& 1.91 & 4.24 & 9.62 \\
4&89.0& 0.23 & 4.01 & 10.39 \\
5&134.3&0.23 & 3.47 & 10.40\\
6&179.0&1.65 & 3.80 & 11.03\\
7&253.3&0.29 & 2.48 & 11.68\\
8&299.9&0.58 & 1.88 & 15.38\\
\hline
&\multicolumn{4}{c}{CAHF (all $M_S$)/CASCI--SO}\\
        \hline
1&0.0 & 0.56 & 1.69 & 14.62 \\
2&28.7& 1.77 & 3.20 & 12.62 \\
3&70.6& 1.95 & 4.23 & 9.64 \\
4&88.8& 0.21 & 4.10 & 10.34\\
5&133.8&0.23 & 3.46 & 10.40\\
6&178.4&1.65 & 3.79 & 11.07\\
7&252.4&0.32 & 2.46 & 11.70\\
8&299.2&0.57 & 1.85 & 15.39\\
\end{tabular*}
\end{table}

\section{Conclusion}

We have investigated the application of the CAS\-SCF/SI--SO {\it ab initio}
method to the calculation of crystal field splitting and magnetic anisotropy in
complexes of trivalent lanthanide ions. The two main ingredients of this method
are: (i) Coupling of Slater determinants into Russell--Saunders-like terms by
configuration interaction in the active space 4f$^n$; (ii) Coupling of those
terms into $J$-like multiplets by spin--orbit state interaction. CASSCF
performs step (i) but uses state-dependent molecular orbitals. This complicates
step (ii) because the SI--SO program has to calculate matrix elements between
states expressed in mutually non-orthogonal orbitals. Based on the fact that
``4f molecular orbitals'' in Ln(III) complexes are almost pure atomic 4f
orbitals we have suggested that significant state-dependence of the CASSCF
molecular orbitals is not expected. This is corroborated by the already common
practice of state-averaging CASSCF over a large number of 4f$^n$ states. If so,
a single set of 4f-configuration-averaged orbitals may be used to represent all
states. As a result, steps (i) and (ii) may be combined in a convenient single
diagonalization on the Slater determinant basis.

\begin{acknowledgments}
W.V.d.H. thanks the University of Melbourne for a McKenzie Postdoctoral
Fellowship. A.S. acknowledges financial support from the Australian Research
Council, through a Discovery Grant, project ID: DP150103254.
\end{acknowledgments}

\bibliography{Ln_confav_rev2} 

\end{document}